\documentclass[aps,preprint,nofootinbib,superscriptaddress,prc]{revtex4-1}
\usepackage{amssymb}

\usepackage{epsfig}
\usepackage[bookmarksnumbered,bookmarksopen,colorlinks,citecolor=blue,linkcolor=blue]{hyperref}

\usepackage{fancyhdr}
\begin{document}

\fancyhead[c]{ Submitted to `Chinese Physics C'}

\title{Production Cross Section of Neutron-Rich Calcium Isotopes in Heavy Ion Collisions}

\author{Donghong Zhang}

 \affiliation{The Key Laboratory of Beam Technology and Material Modification of Ministry of Education,College of Nuclear Science and Technology, Beijing Normal University, Beijing 100875, China}
 \affiliation{Beijing Radiation Center, Beijing 100875, China}

\author{Wenjie Xie}
\affiliation{The Key Laboratory of Beam Technology and Material Modification of Ministry of Education,College of Nuclear Science and Technology, Beijing Normal University, Beijing 100875, China}
\affiliation{Beijing Radiation Center, Beijing 100875, China}
\affiliation{Department of Physics, Yuncheng University, Yuncheng 044000, China}

\author{Jun Su}
\affiliation{Sino-French Institute of Nuclear Engineering $\&$ Technology, Sun Yat-sen University, Zhuhai 519082, China}

\author{Fengshou Zhang}
\email{fszhang@bnu.edu.cn}
\affiliation{The Key Laboratory of Beam Technology and Material Modification of Ministry of Education,College of Nuclear Science and Technology, Beijing Normal University, Beijing 100875, China}
\affiliation{Beijing Radiation Center, Beijing 100875, China}
\affiliation{Center of Theoretical Nuclear Physics, National Laboratory of Heavy Ion Accelerator of Lanzhou, Lanzhou 730000, China}

\begin{abstract}
Based on the isospin-dependent quantum molecular dynamics model along with the GEMINI
model, heavy-ion collisions at intermediate energies are studied. We calculate the production cross sections of different fragments for reactions of $^{112}$Sn+$^{112}$Sn
and $^{124}$Sn+$^{124}$Sn at different beam energies. The species and production cross sections of neutron-rich isotopes are generally dependent on the isospin of the system and the incident energies. The isotopes $^{48}$Ca and $^{54}$Ca are more productive for the neutron-rich system at 30 to 150 MeV/nucleon.
\end{abstract}

\maketitle

\begin{center}
\textbf{I. INTRODUCTION}
\end{center}

The nuclear multifragmentation is an important reaction mechanism in heavy ion collisions\cite{lab1}. General characteristics of the multifragmentation reaction have been observed since the advent of powerful 4$\pi$ detectors\cite{lab2}. It appears that further improvements are related to the study of many observables and the correlations in the multifragment events. Nuclear
fragmentation is a well-established technique to achieve rare isotope beams\cite{lab3}, and plays an important role in
nuclear physics.

Much progress has been made in the production of neutron-rich nuclides in recent years\cite{lab4}. With the establishment of secondary beam facilities,
radioactive beams of nuclei with large neutron or proton excess have provided a terrific opportunity to investigate the isospin-dependence
of heavy ion collision dynamics\cite{lab5,lab6}. Reactions induced by neutron-rich nuclei provide crucial information on
the isospin dependence of the nuclear equation of state\cite{lab7,lab8}. Moreover, the synthesis and study of neutron-rich nuclides can give much information about the properties of the nuclear structure.

Using a radioactive nuclear beam, one can study the properties of nuclei very far from
the $\beta$ stability line and isospin degrees of freedom in nuclear reactions at wide energy ranges for different
projectile-target combinations\cite{lab9,lab10,lab11,lab12}. Investigation of the nuclear landscape close to
the neutron-drip line\cite{lab13} is widely concerned, for the sake of
explaining the evolution of nuclear structure, with increasing neutron-to-proton ratio (N/Z)\cite{lab14} and understanding
vital nucleosynthesis pathways\cite{lab15}. Therefore it is important to investigate the production of neutron-rich isotopes.

Nuclei in the calcium region are attractive to test nuclear models and have been studied both
experimentally and theoretically. Among the calcium
isotopes, $^{40}$Ca and $^{48}$Ca are doubly magic nuclei. In particular, $^{48}$Ca is the lightest doubly magic nucleus with
neutron excess and the published data indicate the high purity
of its doubly closed-shell structure. $^{48}$Ca nucleus has been used in nuclear reactions for the sake of synthesizing superheavy elements and to investigate neutron transfer reactions over a large number of systems\cite{lab16}. It is known that $^{54}$Ca is close to the drip line of Z=20\cite{lab17}. Thus we are greatly interested in studying the production cross sections of $^{48}$Ca and even heavier isotopes of Ca such as $^{54}$Ca in heavy ion collisions.

\begin{center}
\textbf{II. THEORETICAL FRAMEWORK}
\end{center}

With regard to the existing models some are correlated with statistical descriptions based on multi-body phase space
calculations\cite{lab18} whereas others depict the dynamic evolution of systems resulting from collisions between nuclei by molecular
dynamics\cite{lab19} or stochastic mean field methods\cite{lab20}. The empirical parametrization of fragmentation cross sections helps to predict the mass and the charge distributions in heavy-ion reactions \cite{lab21}.
The statistical abrasion ablation model can reproduce the experimental outcomes of heavy-ion collisions (HICs)\cite{lab22,lab23}. The quantum molecular
dynamics (QMD) model includes the information about transport mechanisms\cite{lab24}. The isospin-dependent Boltzmann-Langevin model has been used to investigate the fragment cross sections\cite{lab25,lab26}.

In this paper, we attempt to study the production cross sections in heavy ion collisions by using the IQMD model along with the statistical decay model GEMINI\cite{lab27}. We carry out a systematic study for
multifragmentation of different systems. Several significant fragmentation observables, including the charge
distributions and the production cross sections of some neutron-rich nucleons, are calculated. The production cross sections of the isotopes of He, O and Ca are calculated, especially $^{48}$Ca and $^{54}$Ca.

Generally the process of HICs at intermediate energies may be fractionized into the violent stage
and deexcitation stage. The prefragments form in the violent stage, and the excited prefragments
decay via light-particle emissions, fissions, or complex fragment emissions in the deexcitation stage. The
many-nucleons system is in nonequilibrium in the violent stage, thus it is necessary to adopt the microscopic dynamical
models for prefragment formation. However, at the deexcitation stage of the excited prefragments, the statistical
descriptions are more effective than the dynamical descriptions\cite{lab28}. In our work, the IQMD model is utilized to
simulate the violent stage of the reactions while the GEMINI model is employed to depict the decays of the prefragments.

The IQMD model includes isospin degree of freedom for nucleons. It has been utilized in the analysis of a
large number of observables in HICs at intermediate energies\cite{lab29}. In the IQMD model, the Hamiltonian $H$ is expressed as
\begin{equation}
H=T+U_{Coul}+\int{V}_{nucl}[\rho(\mathbf{r})]d{\mathbf{r}}.
\end{equation}
Here the first term $T$ represents the kinetic energy, the second term
$U_{Coul}$ represents the Coulomb potential energy, and the third term represents the
local nuclear potential energy. Each term of the local potential
energy-density functional ${V_{nucl}}$ in the work means
\[V_{nucl}=V_{Sky}+V_{sur}+V_{sym},\nonumber\\\]
\[V_{Sky}=\frac{\alpha}{2}\frac{{\rho}^{2}}{\rho_0}+\frac{\beta}{\gamma+1}\frac{{\rho}^{\gamma+1}}{{\rho_0}^{\gamma}},\nonumber\\\]
\[V_{sur}=\frac{g_{sur}}{2}\frac{(\nabla\rho)^2}{\rho_0}+\frac{g^{iso}_{sur}}{2}\frac{{(\nabla\rho_{n}-\nabla\rho_{p})}^2}{\rho_0},\nonumber\\\]
\begin{equation}
V_{sym}=\frac{C}{2}\frac{{(\nabla\rho_{n}-\nabla\rho_{p})}^2}{\rho_0}.
\end{equation}
Here $V_{Sky}$, which includes the two-body interaction term and the three-body interaction term, expresses the
saturation properties of nuclear matter. $V_{sur}$ is the surface term to express the surface of finite nuclei. $V_{sym}$
represents the symmetry term, which plays an important role in recurrencing the isospin-dependence effect in the dynamics.
The parameters of the local potential energy-density functional adopted in our work are shown in Table~\ref{tab1}.

\begin{table}
\tabcolsep=1.5pt \caption{The parameters adopted in the present work.}
{\begin{tabular}{@{}ccccccccc@{}}

\hline\hline
%\hline
   $\alpha$ & $\beta$ & $\gamma$ & $g_{sur}$ & $g{^{iso}_{sur}}$ & $C$ & $g_{\tau}$ & $\rho_0$ & $K$ \\
(MeV) & (MeV) & & (MeV${\cdot}fm{^5}$) & (MeV${\cdot}fm^5$) & (MeV) & (MeV) & ($fm^{-3}$) & (MeV)\\
\hline
-358.4003 &  305.2160 & 7/6 & -0.0127 & 120.2460 & -11.4254 & 0.1630 & 33.2551 & 200\\
\hline\hline
\end{tabular}}

\end{table}

The statistical-model GEMINI is extensively applied in performing sequential decays of hot fragments. It permits not
only light-particle evaporation and symmetric fission, but also all possible binary-decay modes. Nuclear masses with shell and pairing corrections are employed. Nuclear level densities were expressed as a Fermi-gas form. The details of GEMINI
can be found in Ref.~\cite{lab27}.

\begin{center}
\textbf{III. RESULTS AND DISCUSSION}
\end{center}

The multiplicities of fragments in the reaction of $^{197}$Au+$^{197}$Au are depicted as symbols in Fig. 1. For comparison, the experimental data have also been
shown in the figure. In the region 3$\leq$Z$\leq$5, the shape of an exponential decrease is shown for the system at 150 MeV/nucleon. The
multiplicities of fragments with Z=1 are large, due to the contributions of $p$. A general
good agreement is observed between the simulated multiplicities and the experimental data, particularly for
intermediate-mass fragments(6$\leq$Z$\leq$10). The distinctions mainly emerge for light fragments (Z=1,2) and
heavy fragments (Z$>$10). The model overestimates the productions of hydrogen but underestimates the productions
of helium. The productions of the heavy fragments are evidently influenced by two factors. The first factor is the
selection of central events. The second one is the parameters adopted in the construction of clusters at the end of the
IQMD simulations. The good agreement between our
calculations and the data proposes that the IQMD+GEMINI model is a potent and reasonable model for calculating the
cross sections.

\begin{figure}\center
\psfig{file=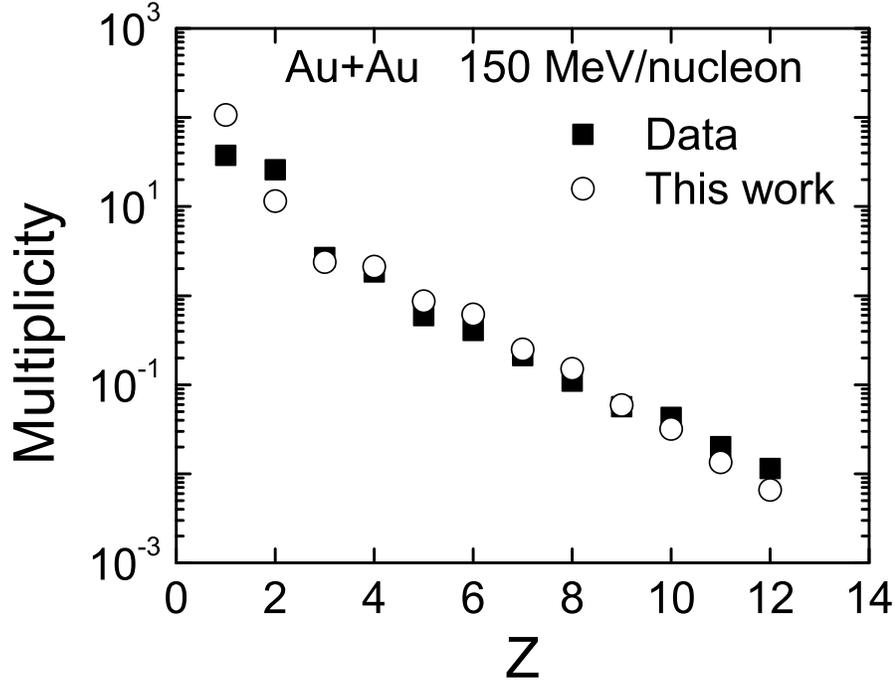,width= 0.8\textwidth}
\caption{Comparison of multiplicities between the present simulations and experimental
data\cite{lab30} for $^{197}$Au+$^{197}$Au central collisions at 150 MeV/nucleon. } \label{aba:1478fig1}
\end{figure}

In order to study the energy and isospin dependence of the charge distributions, we calculate the production cross sections of different fragments for reactions of $^{112}$Sn+$^{112}$Sn and
$^{124}$Sn+$^{124}$Sn at beam energies 50 and 150 MeV/nucleon using the IQMD+GEMINI model. Fig.~\ref{fig2} shows the
calculated production cross sections as a function of fragment charge. It is seen that in the region Z$\leq$40,
there is little difference between the production cross sections of different systems at the same energy. However,
for fragments Z$>$40, the cross sections increase slightly when the neutron of the projectile and target nuclei
become rich. The cross sections in the region Z$\leq$4 increase with the beam energy, but show a decreasing trend
for the heavy fragment(Z$>$4) when the energy is enhanced.

\begin{figure}\center
\psfig{file=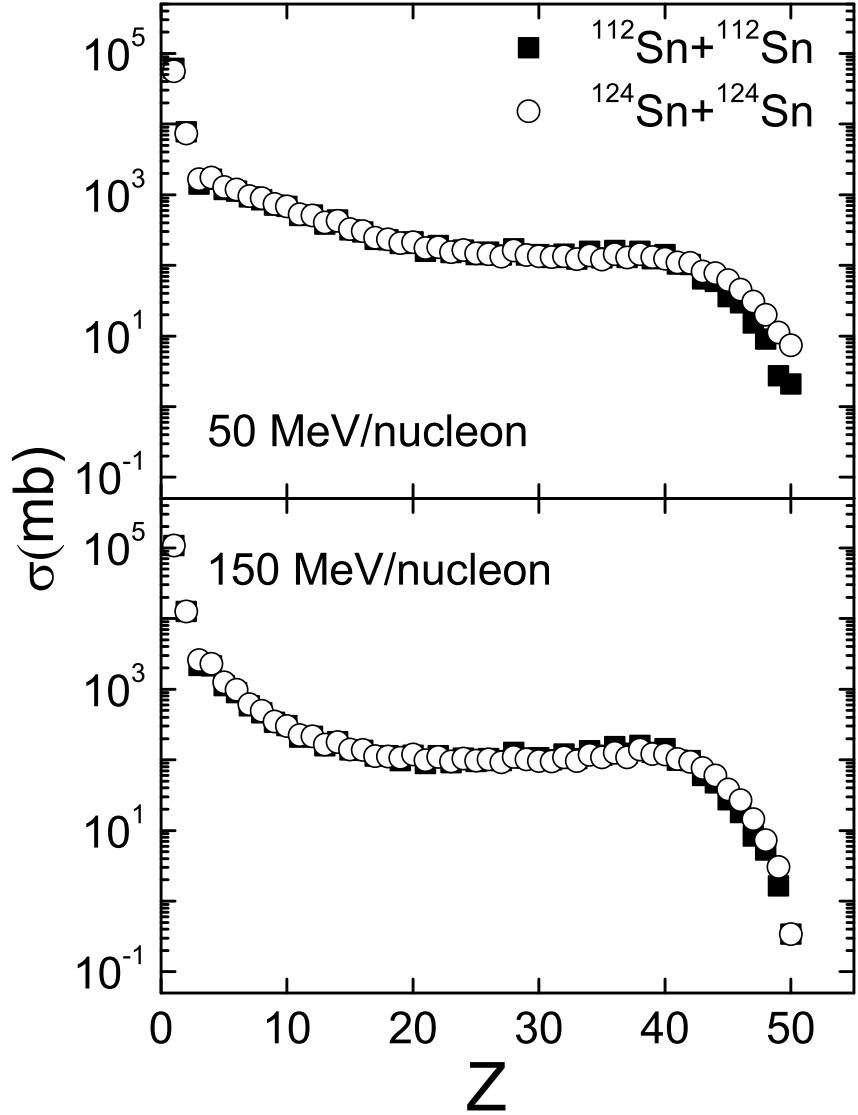,width= 0.8\textwidth}
\caption{Fragment cross sections for the reactions of $^{112}$Sn+$^{112}$Sn(solid squares) and $^{124}$Sn+$^{124}$Sn
(empty circles) at 50 MeV/nucleon and 150 MeV/nucleon. } \label{aba:1478fig2}
\end{figure}

To investigate the production cross sections of different isotopes, we calculated the fragment mass distributions of some nulclides. Fig.~\ref{fig3} displays the fragment cross sections of the system $^{112}$Sn+$^{112}$Sn and $^{124}$Sn+$^{124}$Sn as a
function of fragment mass for isotopes of He, O and Ca at 50 MeV/nucleon and 150 MeV/nucleon. It can be seen that the production cross sections of $^{6-8}$He increase slightly  with the increasing neutron number of the system. The production cross sections of
$^{13-15}$O decrease with increasing the neutron number of the system, while the same quantity rises a little in the region A$>$18. We can see that $^{34-37}$Ca are not produced, but $^{53}$Ca is found in the reaction $^{124}$Sn+$^{124}$Sn at 50 MeV/nucleon. It is shown that the cross section of Ca isotopes larger than A=45 obviously increases as compared to the lighter system. At 150 MeV/nucleon, $^{35,37}$Ca and $^{52,54,56}$Ca disappear in the reaction  $^{112}$Sn+$^{112}$Sn, while
$^{38}$Ca is not produced for $^{124}$Sn+$^{124}$Sn. The production cross sections of heavy isotopes of Ca increase
with the neutron number of the colliding systems, but those of the light ones show opposite behaviors. This phenomenon
is mainly due to the isospin effect on the multifragmentation, since reaction conditions are the same except for the ratios
of neutron to proton. We
can observe that the production cross sections of He increase with the increasing energy, while the fragments of O and Ca show the opposite trend. The peak values appear at $^{4}$He, $^{16}$O and $^{42}$Ca, which are relatively stable isotopes.

\begin{figure}\center
\psfig{file=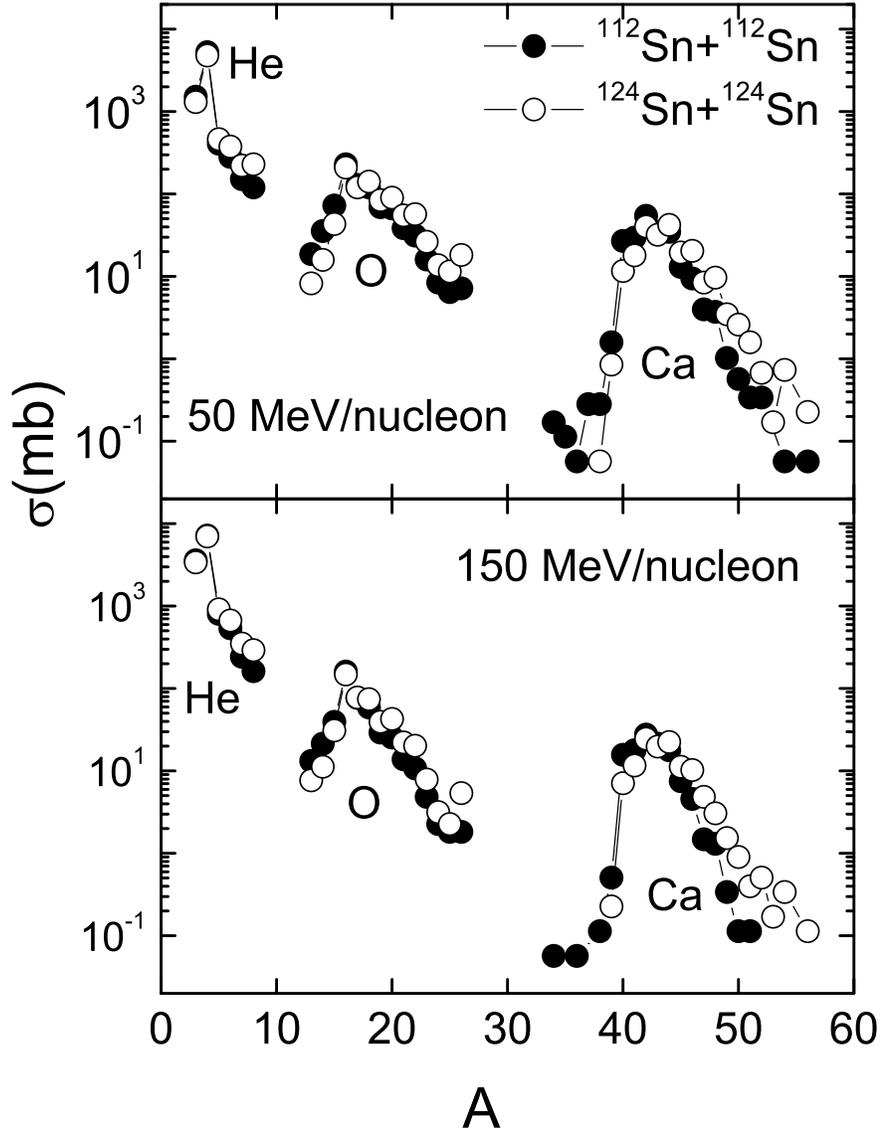,width= 0.8\textwidth}
\caption{Fragment cross sections of He, O and Ca
for the reactions of $^{112}$Sn+$^{112}$Sn(solid circles) and
$^{124}$Sn+$^{124}$Sn(empty circles) at 50 MeV/nucleon and 150 MeV/nucleon. } \label{aba:1478fig3}
\end{figure}

In fact, we are more concerned with the neutron-rich isotopes of Calcium. Fig.~\ref{fig4} and Fig.~\ref{fig5} show the
production cross sections of $^{48}$Ca and $^{54}$Ca as a function of the beam energy for the above-mentioned two reactions. It can be noted that the production cross section of $^{48}$Ca decreases with increasing energy for both systems. The data of $^{54}$Ca show the similar tendency as $^{48}$Ca when the energy changes. The production cross section of $^{48}$Ca and $^{54}$Ca increase obviously with the neutron of the colliding systems for all the beam energies. This phenomenon
is mainly due to the isospin effect on the multifragmentation since other reaction conditions, except the ratios
of neutron to proton, are the same.

\begin{figure}\center
\psfig{file=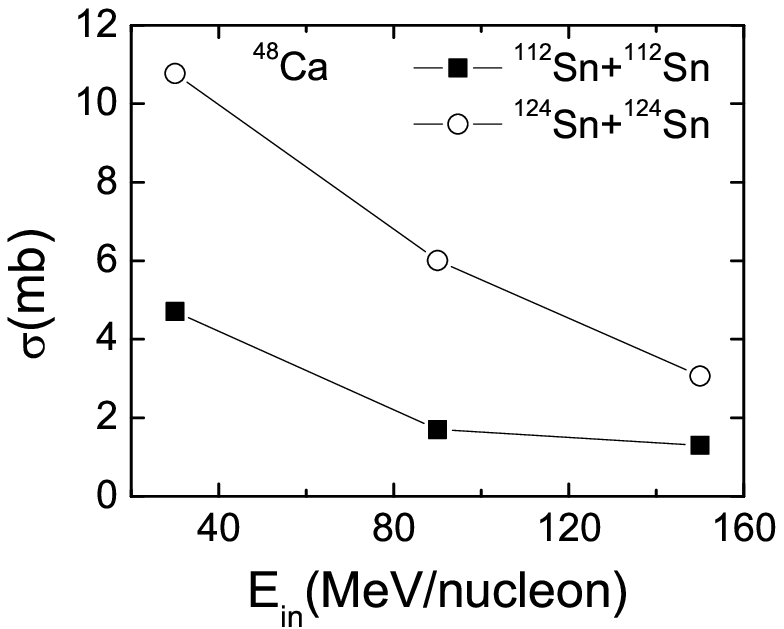,width= 0.8\textwidth}
\caption{Production cross sections of $^{48}$Ca
for the reactions of $^{112}$Sn+$^{112}$Sn(solid squares) and
$^{124}$Sn+$^{124}$Sn(empty circles) from 30 MeV/nucleon to 150 MeV/nucleon. } \label{aba:1478fig4}
\end{figure}

\begin{figure}\center
\psfig{file=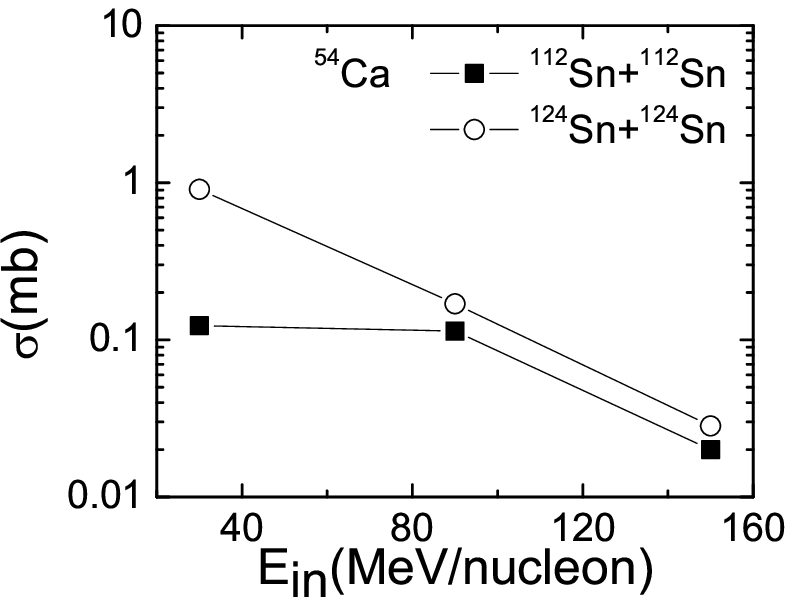,width= 0.8\textwidth}
\caption{Production cross sections of $^{54}$Ca
for the reactions of $^{112}$Sn+$^{112}$Sn(solid squares) and
$^{124}$Sn+$^{124}$Sn(empty circles) from 30 MeV/nucleon to 150 MeV/nucleon. } \label{aba:1478fig5}
\end{figure}

As already calculated, we can observe that the charged-particle cross sections of the systems are almost the same at
the same energy in the small charge area, but display a slight variation for the larger charge. The trends of the production
cross sections as a function of mass for different fragment species are almost from up-sloping to down-sloping, and the
peak mostly corresponds to the stable nucleus, which is easy to understand. Because of the decay of the neutron-rich nucleus,
the yield of the stable nucleus is larger in general. For $^{112}$Sn+$^{112}$Sn, the mass of the projectile and target is
smaller than the other reaction system, and the light isotopes are more produced when considering the fragments of the
same element, which displays a clear isospin effect. The fragment species do not vary for He and O isotopes, but some differences appear when taking Ca into account. The heavy system produces more kinds of heavy isotopes such as $^{48}$Ca and $^{54}$Ca, while the lighter one generates
more light isotopes. The cross sections of He isotopes increase when the energy increases, while those of O and Ca
display opposite changes.

\begin{center}
\textbf{IV. CONCLUSIONS }
\end{center}

In summary, we have studied the fragment cross sections of reactions $^{112}$Sn+$^{112}$Sn and $^{124}$Sn+$^{124}$Sn at
beam energies from 30 MeV/nucleon to 150 MeV/nucleon by using the IQMD+GEMINI model. The results have shown that the production cross
sections of the heavy isotopes such as $^{48}$Ca and $^{54}$Ca for the same element enhance with the increasing neutron-proton ratios of the colliding systems,
which indicates the prominent isospin effect on the process of multifragmentation. From the theoretical simulation, it is
clear to see that neutron-rich nuclei $^{48}$Ca and  $^{54}$Ca are more produced for the neutron-rich system. In most cases, we find more such isotopes when the incident energies decline. These conclusions may serve as good probes for gaining heavy isotopes and further development related to it.

\vspace{-1mm}
\centerline{\rule{80mm}{0.1pt}}
\vspace{2mm}

\begin{center}
\textbf{ACKNOWLEDGEMENTS}
\end{center}

This work was supported by the National Natural Science Foundation of China under Grand Nos. 11025524 and 11161130520, National Basic Research Program of China under Grant No. 2010CB832903, and the European Commission¡¯s 7th Framework Programme (FP7-PEOPLE-2010-IRSES) under Grant Agreement Project No. 269131.

\end{document}